\documentclass[preprint,showpacs,preprintnumbers,amsmath,amssymb,aip]{revtex4-1}

\usepackage{graphicx}
\usepackage{dcolumn}
\usepackage{bm}
\usepackage{psfrag}
\usepackage[T1]{fontenc}

\begin{document}

\preprint{APS/123-QED}

\title{Transition to three dimensional flow in thermal convection with spanwise
rotation}

\author{K. L\"udemann and A. Tilgner}

\affiliation{Institute of Astrophysics and Geophysics, University of G\"ottingen,
Friedrich-Hund-Platz 1, 37077 G\"ottingen, Germany }

\date{\today}

\begin{abstract}
We investigate by direct numerical simulation Rayleigh-B\'enard convection in a
rotating rectangular cell with rotation vector and gravity perpendicular to each
other. The flow is two dimensional near the onset of convection with convection
rolls aligned parallel to the rotation axis of the boundaries. At a sufficiently
large Rayleigh number, the flow becomes unstable to three dimensional
disturbances which changes the scaling of heat transport and kinetic energy with
Rayleigh number. The mechanism leading to the instability is identified as an
elliptical instability. At the transition, the Reynolds and Rossby numbers
$\mathrm{Re}$ and $\mathrm{Ro}$ based on the kinetic energy of the flow are related by
$\mathrm{Re} \propto \mathrm{Ro}^{-2}$ at small $\mathrm{Ro}$ with a geometry
dependent prefactor.
\end{abstract}

\maketitle

\section{Introduction}
Two experiments were recently built which place a thermal convection cell in a
centrifuge. The boundary near the axis of the centrifuge is cooled while the
opposite boundary is heated. The rotation rate of the centrifuge is fast enough
so that the centrifugal acceleration acts as an effective gravity inside the
convection cell and the Earth's gravitational field becomes negligible. The
experiment by \citet{Jiang20} uses liquids as working fluids and relies on a
large rotation rate to generate large Rayleigh numbers. The intention of the
experiment by \citet{Menaut19} on the other hand is to study compressible
convection and uses gas as convecting fluid.

In addition to creating a large effective gravity, the rotation also introduces
a Coriolis acceleration which promotes 2D flow structures in geostrophic
balance. The centrifuged convection cells may be regarded as models of the
equatorial regions of the Earth's core or some deep planetary atmosphere in
which buoyancy drives columnar convection rolls aligned with the rotation axis
of the planet. Strict geostrophic balance requires all terms in the momentum
equation apart from the pressure and Coriolis terms to be negligible. Viscosity
becomes important in Ekman layers adjacent to boundaries perpendicular to the
rotation axis so that the flows in the experiments are never exactly 2D. However,
these deviations from 2D tend to zero for the Ekman number tending to zero so
that we will associate the experimental flows in nearly geostrophic equilibrium
with 2D flows. An important issue in connection with both the planetary
application and the laboratory experiments is to know under which circumstances
the geostrophic equilibrium holds and under which circumstances it is replaced
by 3D flow. The present paper investigates the stability limit of the
geostrophic flow.

It is frequently assumed that geostrophic equilibrium is destroyed as soon as
the Rossby number exceeds some critical value. Nonlinear terms must be
responsible if a flow in geostrophic equilibrium becomes unstable to 3D flows,
and the Rossby number quantifies the ratio of the advection term in the momentum
equation to the Coriolis term. However, it was found that for rotating
convection in a plane layer rotating about an axis perpendicular to the layer,
the Rossby number based on the actual flow velocities does not provide
us with a criterion to predict when the
flow departs from the rotation dominated regime \cite{Schmit09,Schmit10}. In
rotating convection with spanwise rotation, \citet{Jiang20} also find that the
transition away from geostrophic equilibrium is not characterized by a critical
Rossby number independent of other control parameters if the Rossby number is
based on the free fall velocity. These observations motivate us to find an
accurate criterion for the disruption of geostrophic equilibrium for convection
with spanwise rotation.

We will show that the mechanism responsible for the instability of geostrophic
flows in convection with spanwise rotation is an elliptical instability. This
type of instability was already shown to determine the stability of large scale
circulations in convection \cite{Zwirne20} and to limit the size of convection
rolls \cite{Wang20b}. In the present application, we have to deal with a rotating
frame of reference and we have to determine more accurately the properties of
the elliptical instability to account for the observed variations of the
critical Rossby number at the onset of instability of the geostrophic flow.

The next section describes the mathematical formulation on which the numerical simulations in
the third section are based. The fourth section links the numerical results to
the elliptical instability.

\section{The mathematical model}

We consider in a Cartesian coordinate system $x,y,z$ a rectangular cell of size
$d$ along $z$ and size $L$ along $x$ and $y$. The cell rotates about the
$y-$axis at rate $\Omega$. A temperature difference $\Delta T$ is applied to the
boundaries perpendicular to $z$. The cell is filled with a fluid of density
$\rho$, kinematic viscosity $\nu$, thermal diffusivity $\kappa$, thermal
expansion coefficient $\alpha$ and the effective gravitational acceleration in
the cell is $g$. Introducing units of length, time, velocity and temperature as
$d$, $d^2/\kappa$, $\kappa/d$ and $\Delta T$ and assuming the validity of the
Boussinesq approximation, the equation of evolution for the non dimensional
velocity $\bm v(\bm r,t)$, pressure $p(\bm r,t)$ and temperature deviation from
the static profile $\theta(\bm r,t)$ read
\begin{eqnarray}
\frac{\partial}{\partial t} \bm v + (\bm v \cdot \nabla) \bm v 
+ 2 \frac{\mathrm{Pr}}{\mathrm{Ek}} \bm{\hat y} \times \bm v
&=& 
-\nabla p + \mathrm{Pr} \mathrm{Ra} \theta \bm{\hat z} + \mathrm{Pr} \nabla^2 \bm v
\label{eq:NS} \\
\frac{\partial}{\partial t} \theta + \bm v \cdot \nabla \theta -v_z &=& \nabla^2 \theta
\label{eq:Temp} \\
\nabla \cdot \bm v &=& 0
\label{eq:conti} 
\end{eqnarray}
with the control parameters
\begin{equation}
\mathrm{Pr}=\frac{\nu}{\kappa} ~~~,~~~ 
\mathrm{Ra}=\frac{g \alpha \Delta T d^3}{\kappa \nu} ~~~,~~~
\mathrm{Ek}=\frac{\nu}{d^2 \Omega}
\end{equation}
and where hats denote unit vectors.
In an experimental cell in a centrifuge, the effective acceleration $g$ is due
to the centrifugal acceleration and varies in space. This variation can be
reduced by using a cell with a small ratio of cell size $d$ over distance to the
centrifuge axis. We treat $g$ as a constant in our calculations to avoid a
profusion of control parameters. The computational volume is inspired by the
geometry of the experiment by \citet{Menaut19} and we use $d/L=2$ in all simulations.
For consistency, we always select a Prandtl number of 0.7 representative of a gas.

The walls are assumed to be stress free so that flows near the onset of
convection do not contain Ekman layers and are exactly 2D. The boundaries
perpendicular to $z$ have fixed temperature, whereas the other boundaries are
treated as heat insulating walls.

The equations of evolution were approximated for their numerical simulation by
a method optimized for graphical processing units solving the equations for a
weakly compressible fluid \cite{Tanriv11,Tilgne12b} with speed of sound $c$. The
speed of sound was adjusted so that the Mach number $|\bm v|/c$ never exceeded
$0.1$ at any time and any point inside the flow, and the ratio of the time it
takes sound waves to travel across the cell and the rotation period was small.
The latter criterion requires \cite{Tilgne12b} 
$c^2 \gg (\mathrm{Pr} / (2 \pi \mathrm{Ek}) )^2$.
For the simulations presented here, $c^2 \mathrm{Ek}^2 > 0.5$ in all cases.
The numerical code implemented a finite difference method with fourth order
centered finite differences for all terms except the non linear advection terms
which were discretized by a third order upwinding scheme. The simulations were
run on a uniform collocated grid with resolutions of up to $256 \times 256
\times 512$ grid points in 3D and up to $1024 \times 2048$ grid points in 2D.
A third order Runge-Kutta scheme was used for time stepping. Some of the 2D
simulations were also run with the spectral element method implemented in
NEK5000.

\section{The stability limit of convection flows}
Global quantities like the total kinetic energy or the heat transport depend
differently on control parameters in 2D and 3D flows. This is the reason why the
transition from 2D to 3D is important in geophysical applications. It is also
for this reason that the transition is easily detectable in experiments which
cannot access the full velocity field as conveniently as numerical simulations.

We will consider three global averages extracted from the simulations. The first
of these is the heat transport which is also the most important diagnostic in
experiments. The average heat flux, made dimensionless by dividing it by the
heat conducted in the stationary motionless state, is represented by the Nusselt
number $\mathrm{Nu}$ as the average over the boundary $z=0$ of area $A$ of the
normal derivative of the temperature field. If angular brackets 
$\langle ... \rangle$ denote the time average, $\mathrm{Nu}$ is given by
\begin{equation}
\mathrm{Nu} = \langle \frac{1}{A} \int dx \int dy \partial_z T|_{z=0} \rangle .
\end{equation}
The second quantity of interest is the Reynolds number $\mathrm{Re}$ based on
the total kinetic energy defined as
\begin{equation}
\mathrm{Re} = \frac{1}{\mathrm{Pr}} \langle \frac{1}{V} \int dx \int dy \int dz
|\bm v|^2 dV \rangle^{1/2}
\label{eq:Re}
\end{equation}
where $V$ is the total volume of the cell. The factor of $\mathrm{Pr}$ is
necessary because velocity was made non dimensional with the thermal
diffusivity. Finally, the most direct indicator of a transition from 2D to 3D
flow is the anisotropy defined as the ratio of the energy in the velocity
component parallel to the rotation axis and the energy in the components
perpendicular to the axis of rotation:
\begin{equation}
\frac{\langle \frac{1}{V} \int dx \int dy \int dz v_y^2 dV \rangle}
{\langle \frac{1}{V} \int dx \int dy \int dz (v_x^2+v_z^2) dV \rangle}
.
\label{eq:anisotropy}
\end{equation}

\begin{figure}
\includegraphics[width=8cm]{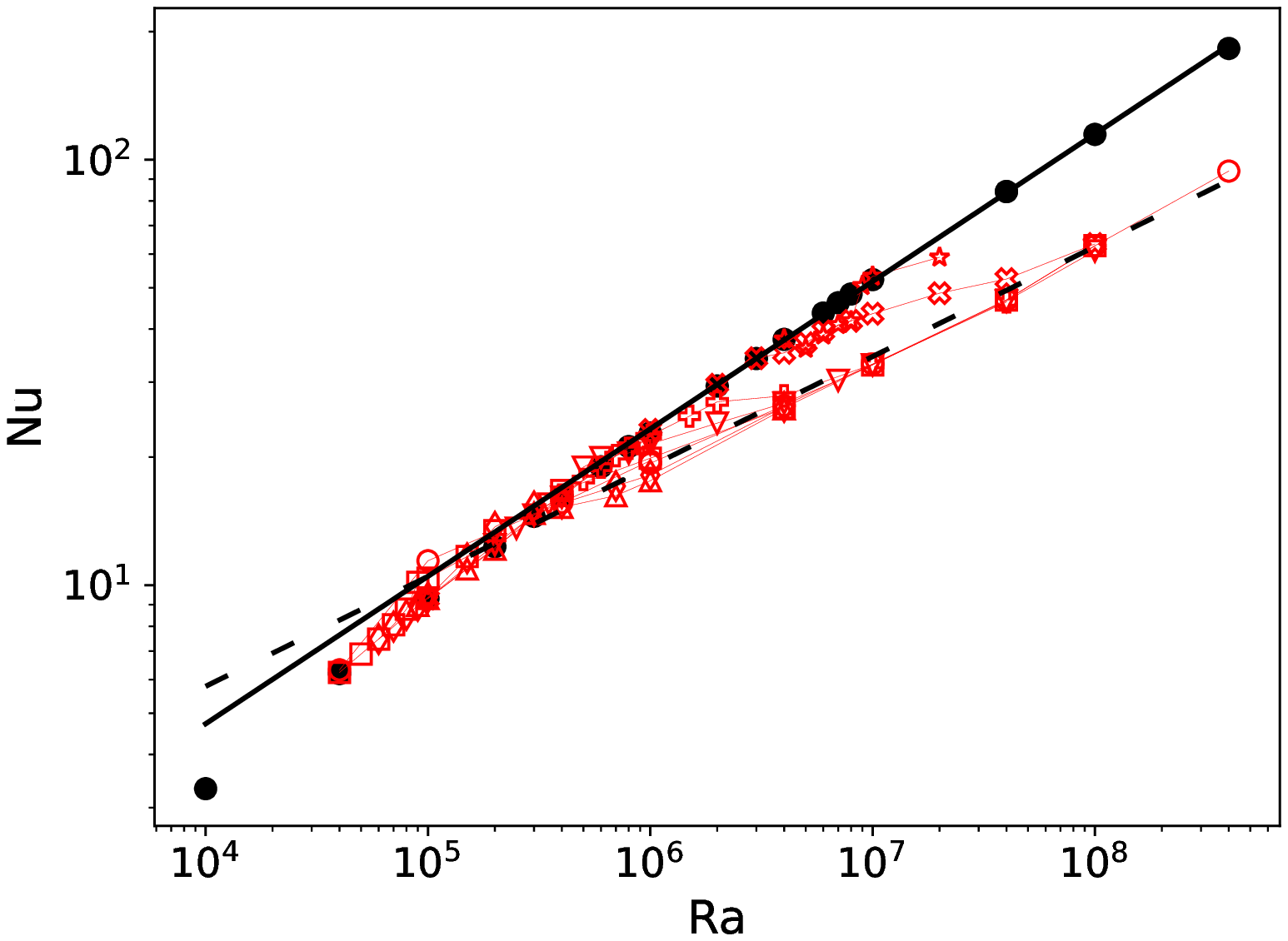}
\includegraphics[width=8cm]{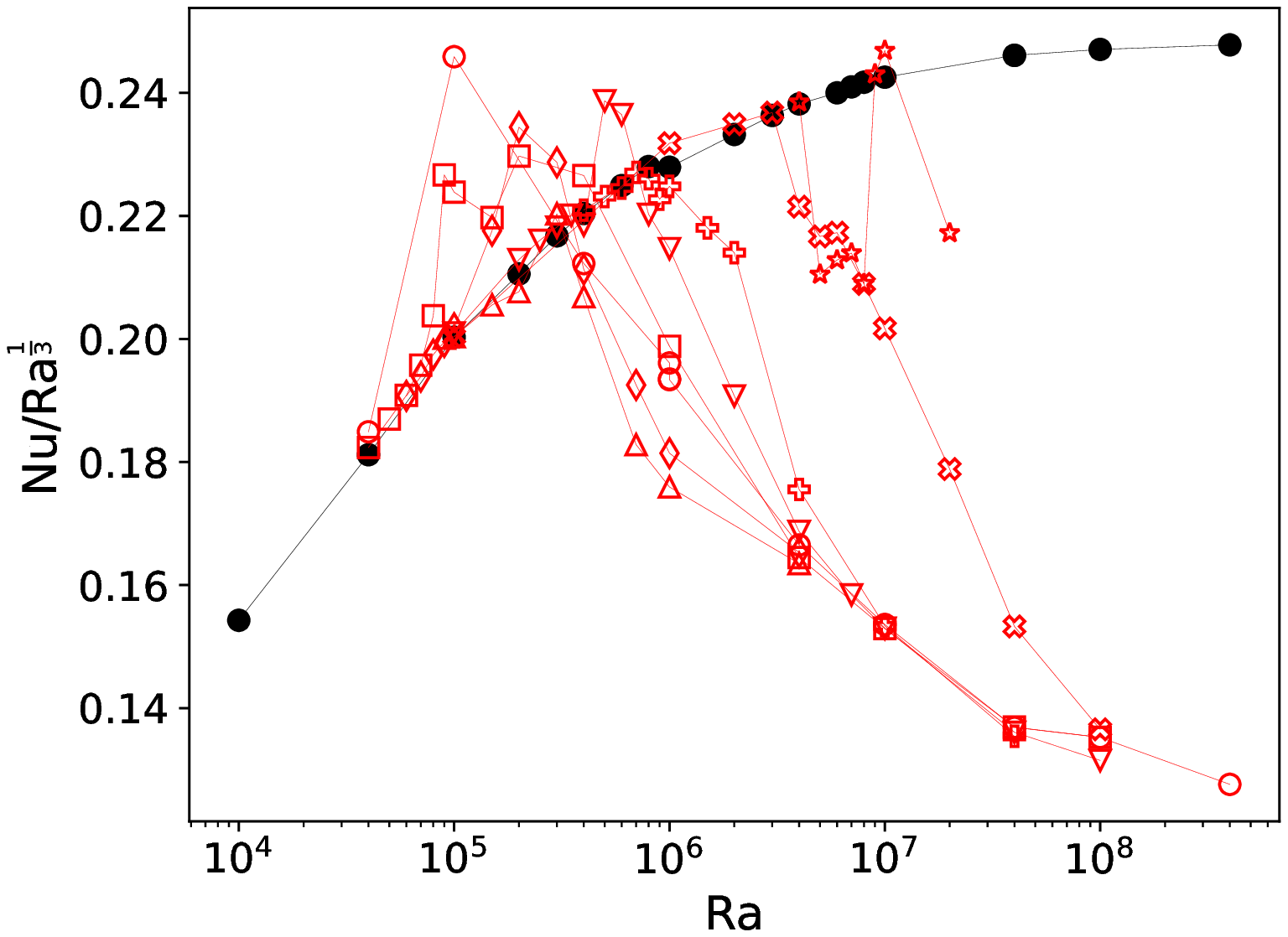}
\caption{
$\mathrm{Nu}$ as a function of $\mathrm{Ra}$ for 2D (black dots) and 3D 
(empty red symbols) simulations for non rotating convection (circles),
$\mathrm{Ek}=10^{-2}$ (squares), $7 \times 10^{-3}$ (diamonds),
$3 \times 10^{-3}$ (triangles up), $10^{-3}$ (triangles down),
$4 \times 10^{-4}$ (plus), $10^{-4}$ (x) and 
$4 \times 10^{-5}$ (stars). The solid and dashed lines in the left panel
indicate the power laws
$\mathrm{Nu}=0.18 \mathrm{Ra}^{0.35}$ and 
$\mathrm{Nu}=0.54 \mathrm{Ra}^{0.26}$, respectively, whereas the right panel shows the
compensated Nusselt number $\mathrm{Nu}/\mathrm{Ra}^{1/3}$.}
\label{fig:1}
\end{figure}

\begin{figure}
\includegraphics[width=8cm]{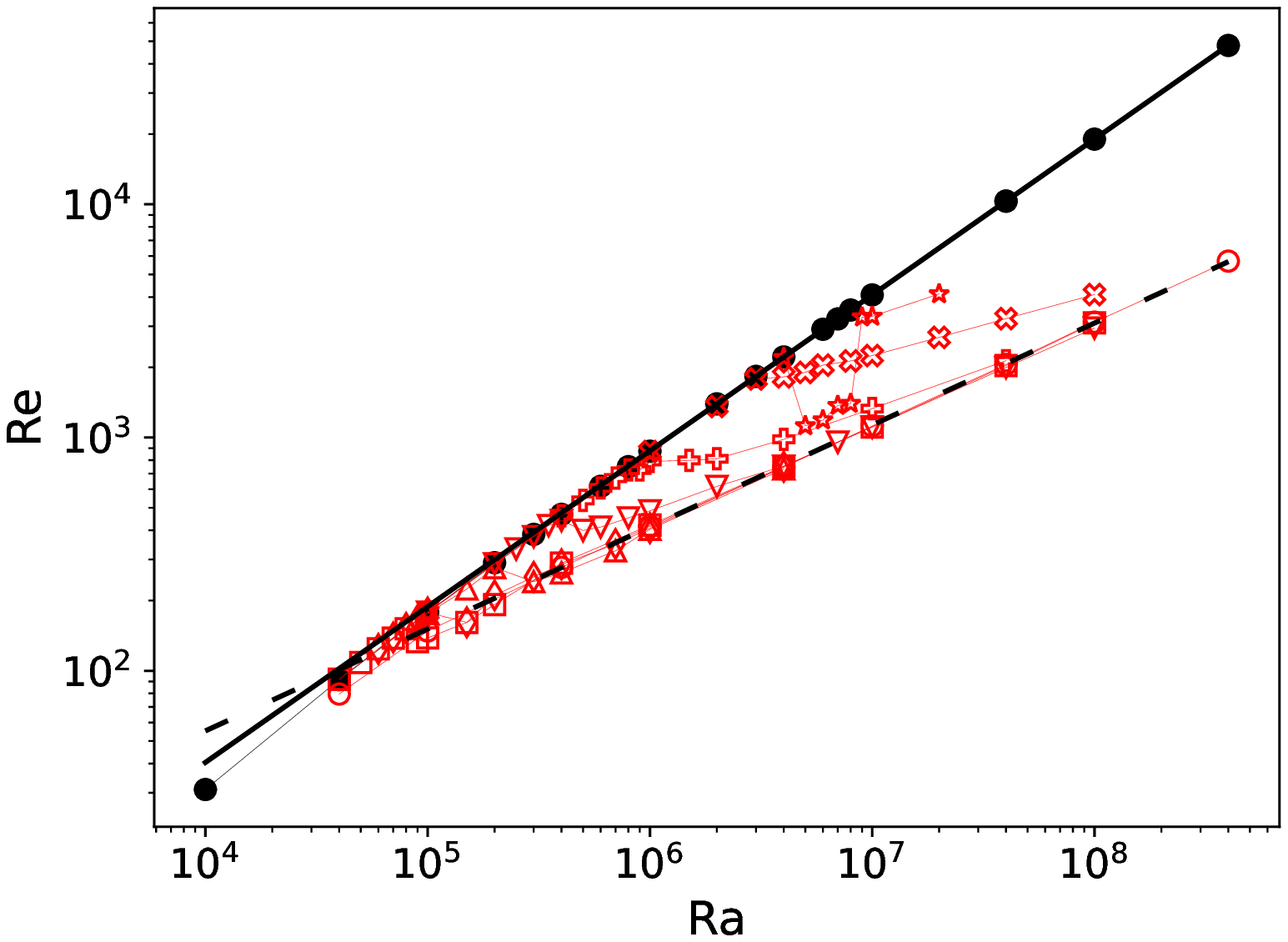}
\includegraphics[width=8cm]{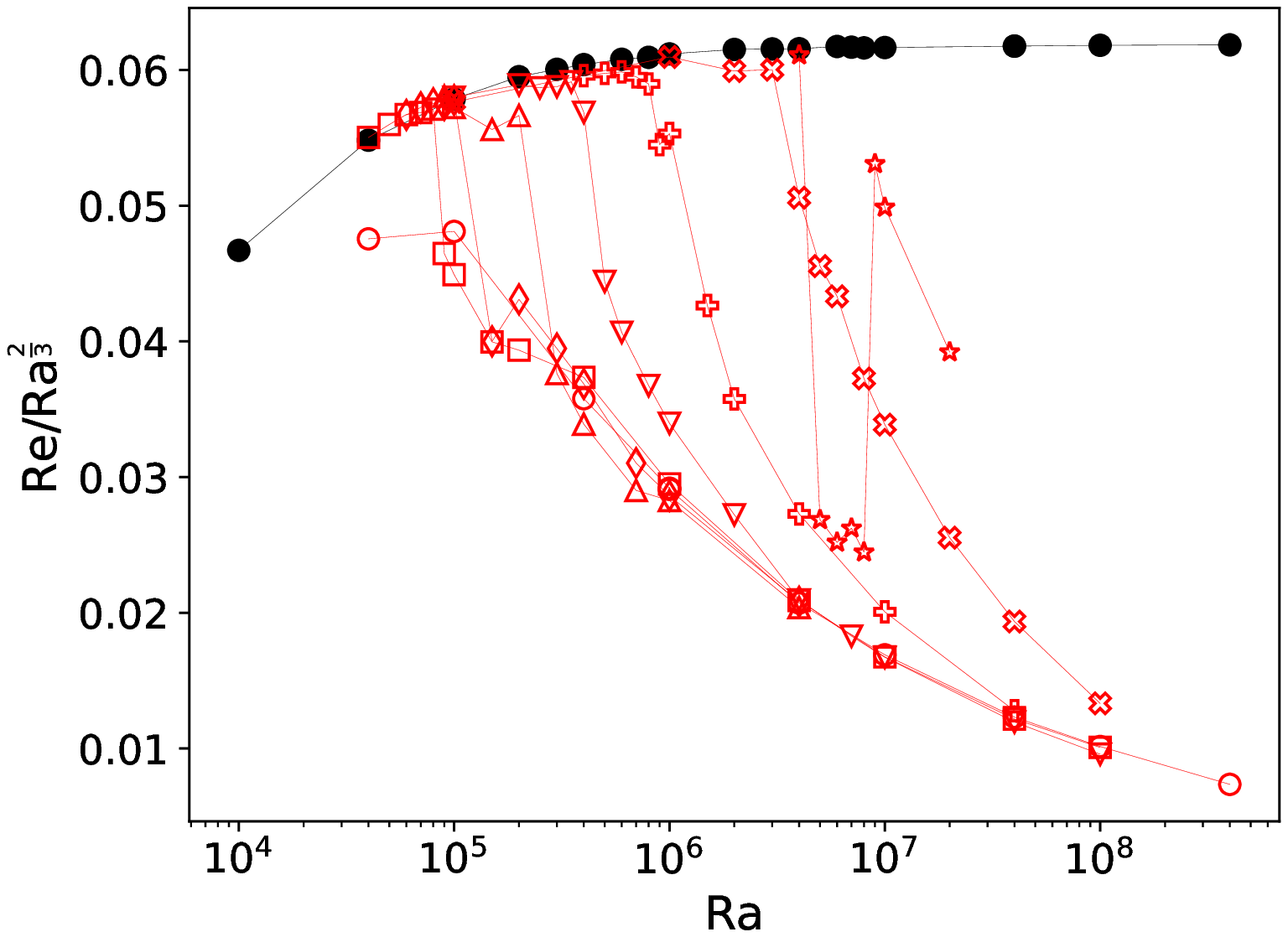}
\caption{
Reynolds number $\mathrm{Re}$ (left panel) and compensated Reynolds number 
$\mathrm{Re}/\mathrm{Ra}^{2/3}$ (right panel) as a function of $\mathrm{Ra}$ with 
the same symbols as in fig. \ref{fig:1}. The solid and dashed lines in the left panel
indicate the power laws
$\mathrm{Re}=0.09 \mathrm{Ra}^{0.67}$ and
$\mathrm{Re}=0.99 \mathrm{Ra}^{0.44}$, respectively.}
\label{fig:2}
\end{figure}

Figs. \ref{fig:1} and \ref{fig:2} show $\mathrm{Nu}$ and $\mathrm{Re}$ for two
types of simulations. The first set of computations simulates 2D flows with
streamlines confined to planes perpendicular to the rotation axis. The Coriolis
term for these flows is conservative and can be absorbed into the pressure
gradient so that these results are independent of $\mathrm{Ek}$. Apart from
lateral boundary conditions, these simulations are identical to older
calculations. \citet{Wen20} investigate steady convection in 2D and find
$\mathrm{Nu}$ and $\mathrm{Re}$ to scale as
$\mathrm{Nu} \propto \mathrm{Ra}^{1/3}$ and
$\mathrm{Re} \propto \mathrm{Ra}^{2/3}$ at constant lateral size of the
convection rolls. For time dependent convection, \citet{vdPoel14} find 
$\mathrm{Nu} \propto \mathrm{Ra}^{2/7}$ and \citet{Wang20} report similar
exponents. The 2D runs in figs. \ref{fig:1} and \ref{fig:2} ended 
after some initial transient in a time independent state consisting of a single
convection roll. Not surprisingly, the best fits
to these results,
\begin{equation}
\mathrm{Nu}=0.18 \mathrm{Ra}^{0.35}  ~~~~,~~~~
\mathrm{Re}=0.09 \mathrm{Ra}^{0.67}
,
\end{equation}
yield exponents compatible with \citet{Wen20}.
As seen in the compensated plots, the exponents of \citet{Wen20} are
approached at large $\mathrm{Ra}$.

The second set of simulations in figs. \ref{fig:1} and \ref{fig:2} is fully 3D
and hence dependent on $\mathrm{Ek}$. One expects the dominant Coriolis term at
low $\mathrm{Ek}$ and $\mathrm{Ra}$ to force the flow into a 2D geometry
identical to the geometry enforced by construction in the first set of
simulations. In that case, both sets of simulations must yield the same
$\mathrm{Nu}$ and $\mathrm{Re}$. As $\mathrm{Ra}$ is increased at constant $\mathrm{Ek}$, the
Coriolis term becomes negligible compared with the advection term and all global
quantities approach the values they have in the non rotating case.
\citet{Pandey16} for instance report for 3D non rotating convection with free slip
boundaries that $\mathrm{Nu} \propto \mathrm{Ra}^{0.27}$. It is straightforward
to compare our simulations with existing results on non rotating convection. The
situation is different for simulations which implement periodic lateral boundary
conditions and allow a mean flow to appear which can significantly modify the
convection \cite{Rouhi21}.

The expected
qualitative behaviour is indeed observed in figs. \ref{fig:1} and \ref{fig:2}
which show series of simulations in which $\mathrm{Ra}$ varies at several fixed
$\mathrm{Ek}$. At each of the $\mathrm{Ek}$, the results are close to the 2D
simulations at low $\mathrm{Ra}$, whereas at large $\mathrm{Ra}$, they asymptote
to the scalings
\begin{equation}
\mathrm{Nu}=0.54 \mathrm{Ra}^{0.26}  ~~~~,~~~~
\mathrm{Re}=0.99 \mathrm{Ra}^{0.44}
\end{equation}
indicative of 3D convection. A smaller $\mathrm{Ek}$ requires a larger
$\mathrm{Ra}$ to start the transition from 2D to 3D. The compensated plot for
$\mathrm{Nu}$ shows a conspicuous overshoot of $\mathrm{Nu}$ at the transition
for the larger Ekman numbers. The origin of this behavior was not investigated
in detail because it disappears at small $\mathrm{Ek}$.

\begin{figure}
\includegraphics[width=8cm]{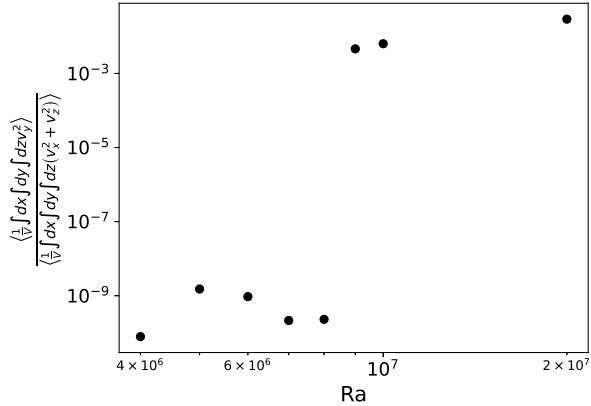}
\caption{
The anisotropy (\ref{eq:anisotropy}) as a function of $\mathrm{Ra}$ for
$\mathrm{Ek}=4 \times 10^{-5}$.}
\label{fig:3}
\end{figure}

\begin{figure}
\includegraphics[width=8cm]{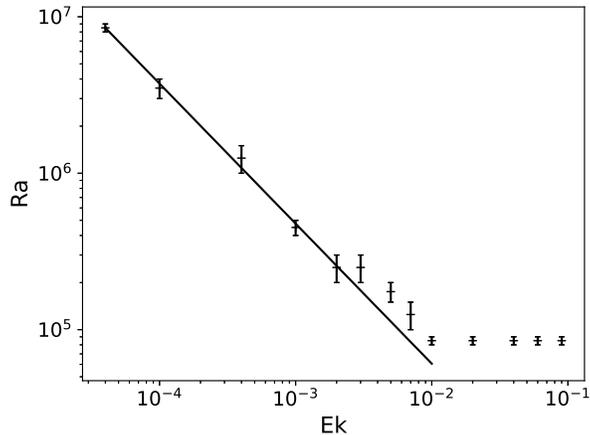}
\caption{
The vertical bars indicate
intervals of $\mathrm{Ra}$ in which the transition from 2D to 3D flows occurs as a 
function of $\mathrm{Ek}$. The solid line is given by
$982 \mathrm{Ek}^{-0.9}$.}
\label{fig:4}
\end{figure}

It is inconvenient to locate the transition from 2D to 3D flows from the above
results because the transition shows as a gradual deviation from a baseline (the
2D results) which is itself contaminated by numerical error. It is much more
reliable to deduce the transition from the anisotropy (\ref{eq:anisotropy})
because this quantity is exactly zero for the 2D flows owing to the free slip
boundaries. Fig. \ref{fig:3} shows as an example the anisotropy as a function of
$\mathrm{Ra}$ for $\mathrm{Ek}=4 \times 10^{-5}$. The numerically evaluated anisotropy at
small $\mathrm{Ra}$ for the supposedly 2D flows is not exactly zero but it is of
the magnitude of round off errors and it is as small as it can be expected to be
in a single precision floating point representation of real numbers. 
Plots like fig. \ref{fig:3} allow us to bracket the interval of $\mathrm{Ra}$ in
which the transition occurs, the lower limit of the interval being the largest
simulated $\mathrm{Ra}$ at which the anisotropy is of
the size of round off errors and the upper limit being the smallest
$\mathrm{Ra}$ at which the anisotropy is clearly different from zero. These
intervals are shown as a function of $\mathrm{Ek}$ in fig. \ref{fig:4}. At large
$\mathrm{Ek}$, rotation is irrelevant and the transitional $\mathrm{Ra}$ is
independent of $\mathrm{Ek}$, whereas for small $\mathrm{Ek}$, the transition
occurs approximately at a Rayleigh number given by $982 \mathrm{Ek}^{-0.9}$.

This observation leads to the hypothesis that buoyancy forces are irrelevant for
the transition at low $\mathrm{Ek}$. One can compare the
characteristic time scales of the dynamics induced by rotation and buoyancy. The
time scale for rotation, which manifests itself for example in the dispersion
relation of inertial waves, is of course the inverse of the rotation rate
$1/\Omega$. A characteristic time scale for buoyancy is the transit time of a
parcel of fluid traversing the cell at free fall velocity, 
$[d/(g \alpha \Delta T)]^{1/2}$. The ratio of both,
$[g \alpha \Delta T / d]^{1/2} /\Omega = (\mathrm{Ra} \mathrm{Ek}^2 / \mathrm{Pr})^{1/2} $,
which may also be identified as a Rossby number \cite{Zhong09,Horn15},
is equal to $0.82$ at $\mathrm{Ek}=10^{-3}$ and equal to $0.066$ at 
$\mathrm{Ek}=10^{-5}$ at the transition. If this ratio is small, the motion of a
fluid parcel changes under the influence of rotational effects on a time scale
on which buoyancy is unable to modify its motion. Buoyancy therefore becomes
irrelevant to the dynamics.

\begin{figure}
\includegraphics[width=8cm]{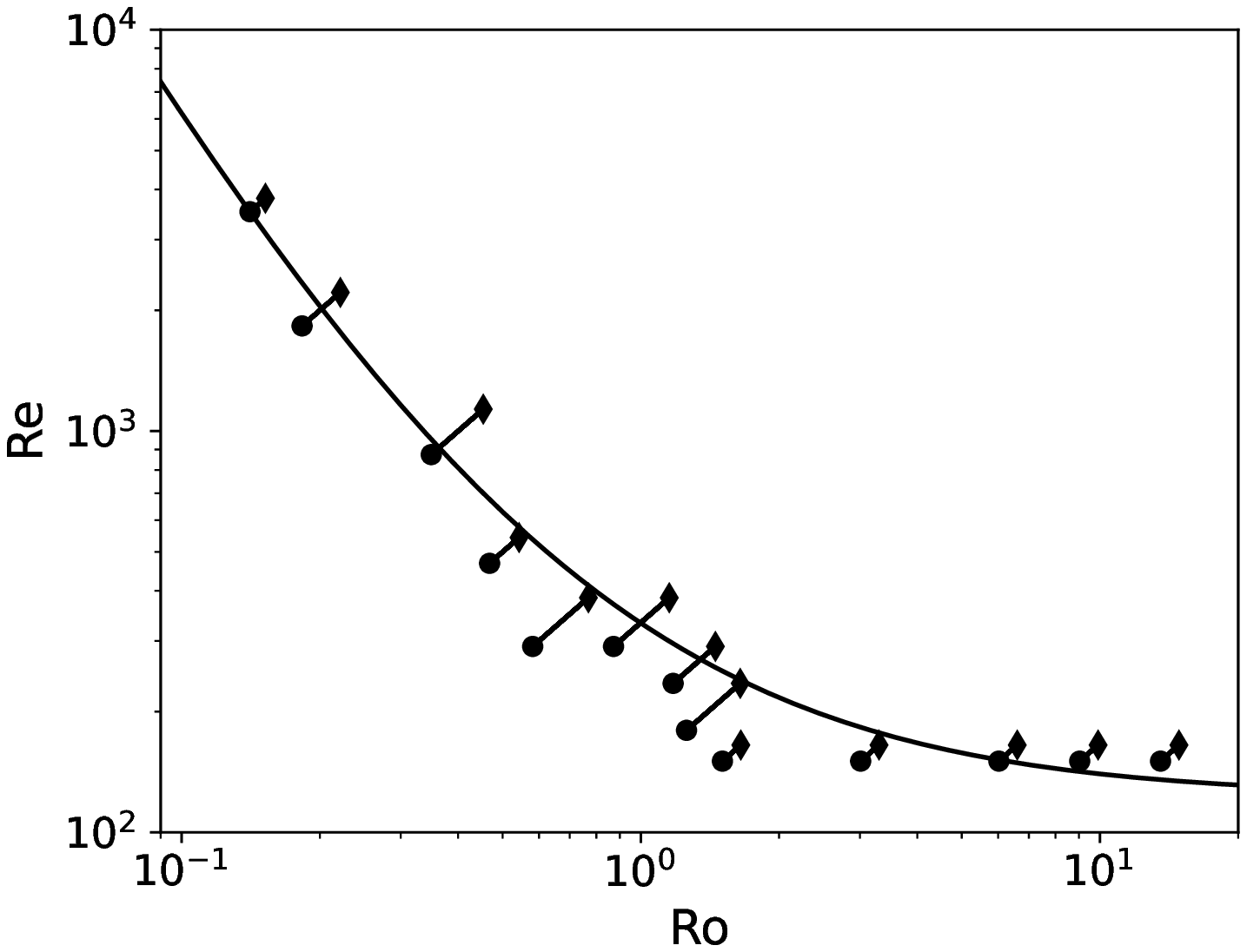}
\includegraphics[width=8cm]{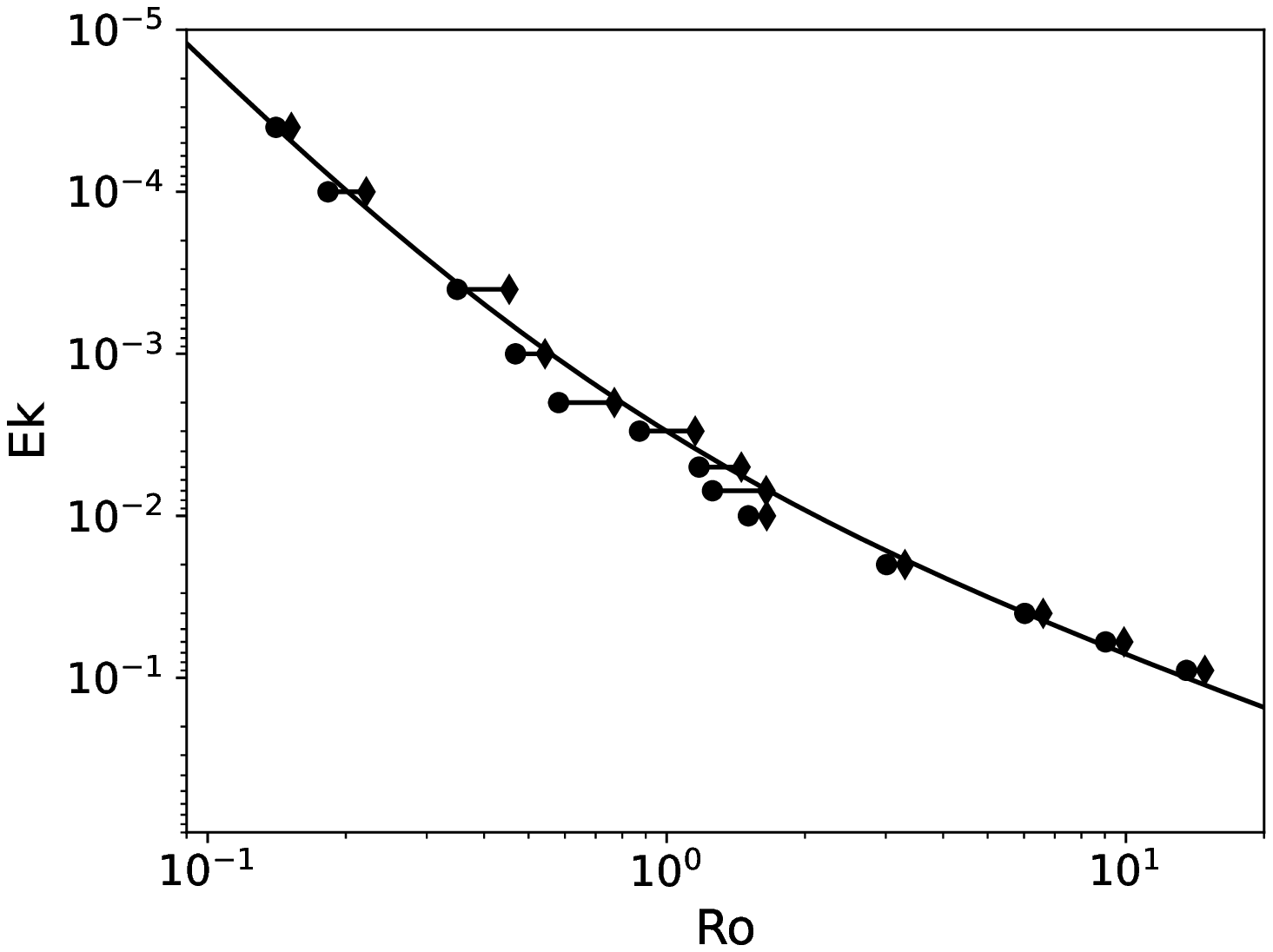}
\caption{
Line segments crossing the stability limit of 2D flows in the 
($\mathrm{Re}$,$\mathrm{Ro}$)-plane (left panel). The continuous curve results from the
stability analysis for an elliptical vortex with $\beta=0.78$ and
$k_\parallel = 2\pi$, $k_\perp =\pi$. The right panel shows the same data and the same
curve in the ($\mathrm{Ek}$,$\mathrm{Ro}$)-plane.}
\label{fig:5}
\end{figure}

This motivates us to describe the transition with parameters independent of
buoyancy. In isothermal flows, the transition from flows patterned by the
Coriolis force to flows independent of rotation is usually thought to be
governed by the Rossby number $\mathrm{Ro}$ defined by 
$\mathrm{Ro} = \mathrm{Ek} \mathrm{Re}$. The intervals of $\mathrm{Ra}$ in fig. \ref{fig:4}
containing the transition are translated into intervals of $\mathrm{Re}$ by
simply reading off the $\mathrm{Re}$ for the $\mathrm{Ra}$ at the lower end of
the interval from the 2D simulations, and by computing a Reynolds number by
extrapolating the $\mathrm{Re}(\mathrm{Ra})$ dependence of 2D flows to the
Rayleigh number at the upper end of the intervals. These two Reynolds numbers,
together with the two Rossby numbers computed from $\mathrm{Ro} = \mathrm{Ek}
\mathrm{Re}$, form endpoints of line segments in the
($\mathrm{Re}$,$\mathrm{Ro}$)-plane which must be crossed by the stability limit
of the 2D flows. These line segments are shown in the
($\mathrm{Re}$,$\mathrm{Ro}$)-plane in fig. \ref{fig:5}. For an alternative point of
view, fig. \ref{fig:5} also shows the stability limit in the
($\mathrm{Ek}$,$\mathrm{Ro}$)-plane. One can see that there
is no unique critical Rossby number for the transition. The next section will
show that the stability diagram in the ($\mathrm{Re}$,$\mathrm{Ro}$)-plane can
be explained by assuming that elliptical instability of the convection rolls is
responsible for the transition.

\section{The elliptical instability}
A linear stability analysis of a convection roll in its rectangular cell 
requires a numerical effort comparable to the simulation of the original convection
problem itself. It is more tractable and at the same time more illuminating to
study the linear stability of a simplified flow using an approximate treatment 
of the boundaries.

As a simple model of the 2D laminar convection flow, we will consider a flow
with elliptical streamlines of the form
$((\gamma-\epsilon)z,0,(-\gamma-\epsilon)x)$ 
in a frame of reference rotating at rate $\Omega$ about the $y-$axis. The
constants $\gamma$ and $\epsilon$ represent in dimensional variables
a spatially uniform vorticity divided
by 2 and a rate of strain, respectively. The streamlines of this flow are ellipses
in the $(x,z)-$plane. The ratio of the large and small major axes of the
ellipses is
$\alpha=\sqrt{(1+\beta)/(1-\beta)}$ 
with the definition $\beta=\epsilon/\gamma$.
The calculation in this section parallels earlier work \cite{Bayly86,
Landma87,Craik89,Kerswe02} except for the treatment of the boundaries. 

We cannot treat the model with the same non dimensional variables as in the
previous section because the thermal diffusivity is absent and there is no
inherent length scale. We choose $1/\gamma$ as time scale and express all
lengths in terms of $d$ which is arbitrary at present and which we will later
identify with the size of the convection cell along the direction of gravity.
Velocities are then given in multiples of $\gamma d$. Switching now to non
dimensional variables, the elliptical base flow $\bm U$ is given by
\begin{equation}
\bm U = \bm A \bm r ~~~,~~~
\bm A =
\begin{pmatrix}
0  &  0  &  1-\beta  \\
0  &  0  &  0  \\
-1-\beta  &  0 &  0  
\end{pmatrix}
.
\label{eq:base_flow}
\end{equation}
The full velocity field $\bm u$ is represented as the sum $\bm u = \bm U + \bm
u'$ of the base flow and the perturbation $\bm u'$ which has to satisfy
\begin{eqnarray}
\frac{\partial}{\partial t} \bm u' + (\bm U \cdot \nabla) \bm u' + (\bm u' \cdot \nabla) \bm U
+ (\bm u' \cdot \nabla) \bm u'
+ 2 \omega \bm{\hat y} \times \bm u' 
&=& 
-\nabla p' + \eta \nabla^2 \bm u' 
\label{eq:u'} \\
\nabla \cdot \bm u' &=& 0
\end{eqnarray}
where $p'$ is a pressure variable, $\omega= \Omega/\gamma$ is the non
dimensional rotation rate of the frame of reference, and
$\eta = \nu / (\gamma d^2)$ quantifies viscous dissipation.

Eq. (\ref{eq:u'}) is solved exactly by velocity and pressure fields of the form 
$\bm u'(\bm r,t) = \bm u_0(t) e^{i \bm k(t) \cdot \bm r}$,
$p(\bm r,t) = p_0(t) e^{i \bm k(t) \cdot \bm r}$
provided that the time dependent amplitude $\bm u_0(t)$ and
the wavevector $\bm k(t)$ obey
\begin{eqnarray}
\frac{d}{d t} \bm u_0 + \bm A \bm u_0 + 2 \bm \omega \bm u_0
-2 \frac{\bm k \bm k^T}{|\bm k|^2} (\bm A + \bm \omega) \bm u_0 + \eta |\bm k|^2 \bm u_0
&=& 
0
\label{eq:u0} \\
\frac{d}{d t} \bm k + \bm A^T \bm k &=& 0
\label{eq:k}
\end{eqnarray}
with the definition
\begin{equation}
\bm \omega=
\begin{pmatrix}
0  &  0  &  \omega \\
0  &  0  &  0  \\
-\omega  &  0  &  0  
\end{pmatrix}
.
\end{equation}
It follows directly from eq. (\ref{eq:k}) that the $y-$component of $\bm k$ is
independent of time and that the vector $\bm k$ precesses elliptically about the
$y-$axis. One can thus assume without loss of generality initial conditions for
$\bm k$ of the form
\begin{equation}
\bm k(t=0) = k_0 \begin{pmatrix} 0 \\ \cos \theta \\ \sin \theta \end{pmatrix}
\end{equation}
with arbitrary $k_0$ and $\theta$. At later times, the wavevector is determined
by eq. (\ref{eq:k}) to be
\begin{equation}
\bm k(t) = k_0 
\begin{pmatrix} 
\alpha \sin \theta \sin (\sqrt{1-\beta^2} t) \\ 
\cos \theta \\ 
\sin \theta \cos (\sqrt{1-\beta^2} t) 
\end{pmatrix}
.
\label{eq:k(t)}
\end{equation}
This expression for $\bm k$ inserted into eq. (\ref{eq:u0}) leads to a linear
ordinary differential equation for $\bm u_0$ with time periodic coefficients.
This is a problem which can be reduced to an algebraic eigenvalue problem
following the method detailed in refs. \cite{Bayly86, Landma87,Craik89} of which
we provide a succinct summary for completeness: In a first step, the viscous term
is removed from the equations by introducing a new variable 
$\tilde{\bm u_0}$ as
\begin{equation}
\tilde{\bm u_0} = e^{\eta \int_0^t |\bm k|^2 dt} \bm u_0
\end{equation}
which transforms eq. (\ref{eq:u0}) into
\begin{equation}
\frac{d}{d t} \tilde{\bm u_0} = \bm Q(t) \tilde{\bm u_0}
~~~,~~~
\bm Q(t) = 2 \frac{\bm k \bm k^T}{|\bm k|^2} (\bm A + \bm \omega) -
(\bm A + 2 \bm \omega)
.
\end{equation}
The general solution of this equation may be deduced from the Floquet problem
\begin{equation}
\frac{d}{d t} \bm M = \bm Q(t) \bm M ~~~,~~~ \bm M(t=0)=\bf{1}
,
\label{eq:M}
\end{equation}
the general solution for $\tilde{\bm u_0}$ being a linear combination of
functions of the form
\begin{equation}
\tilde{\bm u_0}(t) = e^{\sigma_i t} \bm f_i(t)
\end{equation}
where the $\bm f_i(t)$, $i=1,2,3$ are periodic functions with period $2\pi/\sqrt{1-\beta^2}$ and the
$\sigma_i$ are given by
\begin{equation}
\sigma_i = \frac{\sqrt{1-\beta^2}}{2 \pi} \ln \mu_i
\end{equation}
where the $\mu_i$ are the three eigenvalues of $\bm M$ at $t=2\pi /
\sqrt{1-\beta^2}$.
One can show that one of the eigenvalues equals 1, say $\mu_3=1$, and the other
two eigenvalues obey $\mu_1 \mu_2 = 1$. If there are growing solutions, one has
$\mu_2 < 1 < \mu_1$ and the fastest growing mode has the growth rate
$\sigma_1 = \frac{\sqrt{1-\beta^2}}{2 \pi} \ln \mu_1$.
We now reinstate viscous dissipation with the help of the integral
\begin{equation}
-\eta \int_0^{2\pi/\sqrt{1-\beta^2}} |\bm k|^2 dt 
=
-\eta k_0^2 \left( 1 + \frac{1}{2} (\alpha^2-1) \sin^2 \theta \right)
.
\end{equation}
The growth rate $\sigma$ of the fastest growing solution $\bm u_0$ of eq. (\ref{eq:u0})
is therefore
\begin{equation}
\sigma = \frac{1}{2\pi} \left( \frac{\sqrt{1-\beta^2}}{2 \pi} \ln \mu_1
- 2 \pi \eta k_0^2 \frac{1-\beta \cos^2 \theta}{1-\beta} \right)
.
\label{eq:sigma}
\end{equation}
This equation yields a growth rate for any given $\omega$, $\eta$, $\beta$ and
$k_0$ after a numerical time integration of eq. (\ref{eq:M}) up to $t=2\pi /
\sqrt{1-\beta^2}$ followed by a numerical evaluation of the eigenvalues of 
$\bm M$. 

At this stage of the calculation, 
we make a different use of eq. (\ref{eq:sigma}) than previous papers
which derived this relation. It is now possible to identify control parameters which yield
solutions growing in time and to find for any fixed $\beta$ and $\omega$ the
largest $\eta$ that still allows for growing solutions by optimizing over $k_0$
and $\theta$. However, in order to reproduce the results of the time
integrations of the previous section, it is necessary to model the effect of the
boundaries. This is done by restricting $k_0$ and $\theta$ to combinations which
satisfy
\begin{equation}
k_0 \cos \theta \ge k_\parallel      ~~~,~~~
k_0 \sin \theta \ge k_\perp
.
\label{eq:limit_k}
\end{equation}
The rationale behind this condition is that the normal velocity at the
boundaries of the cell must be zero and that therefore at least half a
wavelength must fit into the cell. This condition applied to the direction along
the rotation axis leads to $k_\parallel = 2 \pi$. Likewise, the components of
$\bm k$ perpendicular to the axis of rotation, taking into account the factor
$\alpha$ in the $x-$component of (\ref{eq:k(t)}), must be larger than $\pi$ to
fit half a wavelength into a cell of the geometry chosen for the direct
simulations, so that $k_0$ must also satisfy $k_0 \sin \theta \ge \pi$
which implies $k_\perp=\pi$. The conditions (\ref{eq:limit_k}) are not exactly
equivalent to the free slip boundaries of the simulations, but they suffice to
represent the geometric constraints imposed by the finite size of the convection
cell as we shall see.

If we apply the defintion (\ref{eq:Re}) for the Reynolds number to the
elliptical flow taking into account that the scales for the non dimensional
velocities are different, we compute
\begin{equation}
4 \int_{-1/4}^{1/4} dx \int_{-1/4}^{1/4} dy \int_{-1/2}^{1/2} dz
\left[ (1-\beta)^2 z^2 + (1+\beta)^2 x^2 \right]
=
\frac{1}{12} \left[ (1-\beta)^2 + \frac{1}{4} (1+\beta)^2 \right]
\end{equation}
and conclude that we have to compare a simulation at a certain $\mathrm{Re}$ and
$\mathrm{Ro}$ with a purely elliptical flow parameterized with $\eta$, $\omega$ and
$\beta$ such that
\begin{equation}
\mathrm{Re}=\frac{1}{\eta} 
\sqrt{\frac{1}{12} \left[ (1-\beta)^2 + \frac{1}{4} (1+\beta)^2 \right]}
\end{equation}
and
\begin{equation}
\mathrm{Ro}=\frac{1}{\omega} 
\sqrt{\frac{1}{12} \left[ (1-\beta)^2 + \frac{1}{4} (1+\beta)^2 \right]}
.
\end{equation}

It remains to select a $\beta$ to completely specify the stability limit. Fig.
\ref{fig:5} shows the stability limit in the $(\mathrm{Re},\mathrm{Ro})-$plane
for $\beta = 0.78$ which is already a reasonable fit to the simulation data
considering all the simplifications introduced into the model of purely
elliptical streamlines.

\begin{figure}
\includegraphics[width=8cm]{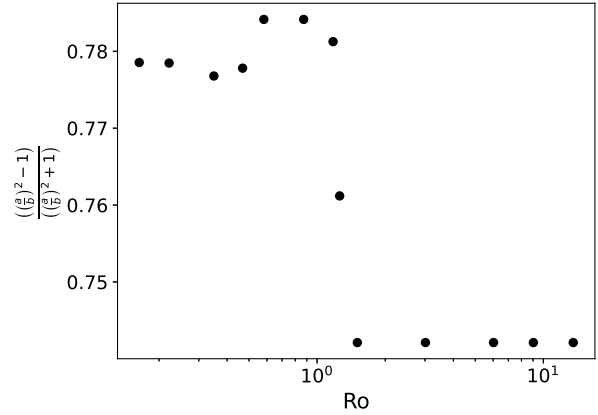}
\caption{
Ellipticities of the streamlines of 2D convection flows at the center of the cell
deduced from the eigenvalues $1/a$ and $1/b$ of the Jacobian of the velocity field.
Ellipticity is defined as the parameter $\beta$ in eq. (\ref{eq:base_flow}).}
\label{fig:6}
\end{figure}

\begin{figure}
\includegraphics[width=5cm]{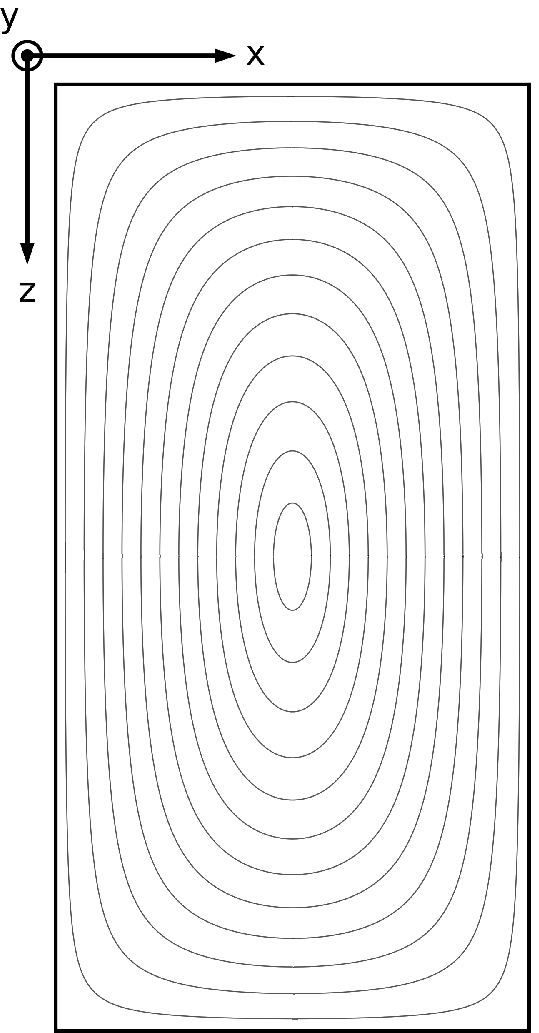}
\includegraphics[width=5cm]{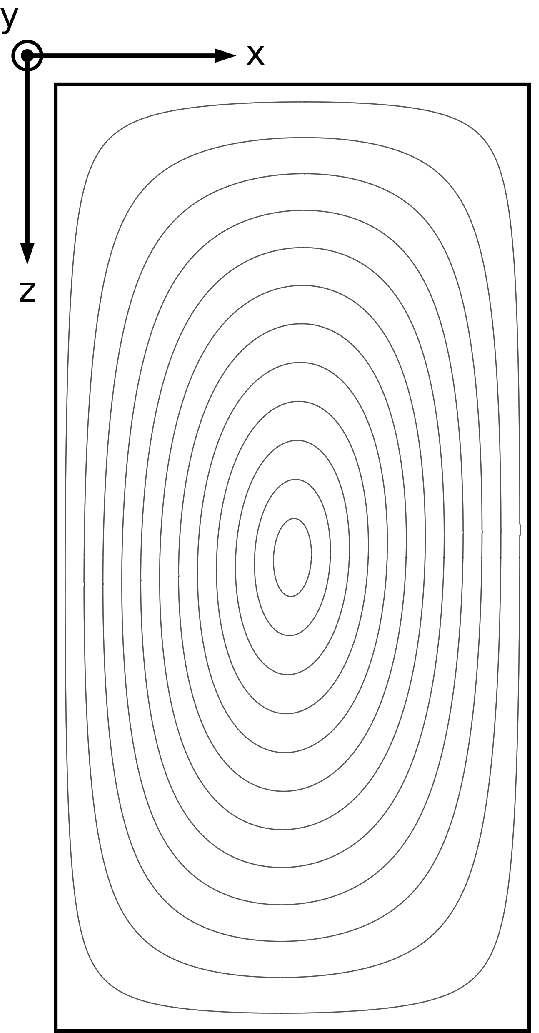}
\includegraphics[width=5cm]{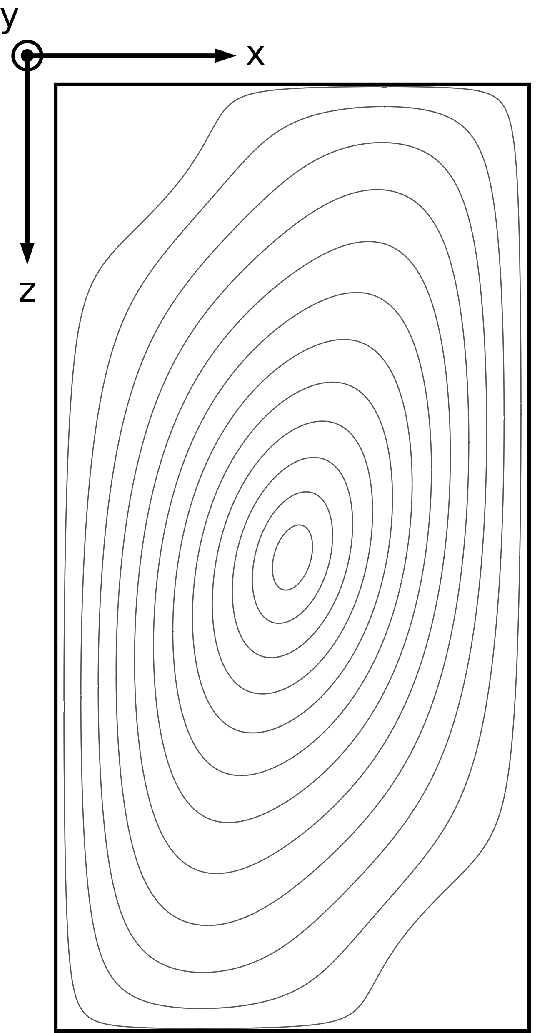}
\caption{
Streamlines of the two dimensional flows at $\mathrm{Ek}=4 \times 10^{-5}$ and 
$\mathrm{Ra}=8 \times 10^6$ corresponding to $\mathrm{Ro}=6.6 \times 10^{-2}$ (left panel), 
$\mathrm{Ek}=7 \times 10^{-3}$ and $\mathrm{Ra}= 10^5$ corresponding to
$\mathrm{Ro}=1.24$ (middle panel) and
$\mathrm{Ek}=9 \times 10^{-2}$ and $\mathrm{Ra}= 8 \times 10^4$ corresponding to
$\mathrm{Ro}=13.3$ (right panel).}
\label{fig:7}
\end{figure}

A simple estimate of $\beta$ in which the major axes of the ellipse are set
equal to the length of the sides of the convection cell yields $\beta=0.6$.
While the streamlines of the convection roll espouse the shape of the cell near
the boundaries of the cell, the ellipticity of these streamlines near the center
of the cell is uncertain. We therefore compute the eigenvalues of the Jacobian
of the velocity field at the center of the cell for the 2D flows next to the
transition. For exactly elliptical streamlines, these eigenvalues are the
inverse of the semi major axes. One of the three eigenvalues is zero in 2D
flows. If we call the smaller of the remaining eigenvalues $1/a$ and the last
one $1/b$, we can compute an approximate $\beta$ as
$\left[ (a/b)^2-1 \right] / \left[ (a/b)^2+1 \right]$.
This combination is shown in fig. \ref{fig:6} as a function of $\mathrm{Ro}$.
The value changes abruptly near $\mathrm{Ro} \approx 1$. The change is
relatively small in absolute value and barely visible in plots of streamlines
(see fig. \ref{fig:7}).
The 2D flow is of course insensitive to rotation and the change in the flow
pattern is due to the different Rayleigh  numbers for the flows close to the
transition. Even though the changes in the surrogate $\beta$ are small, its 
systematic variation with $\mathrm{Ro}$ justifies the use of two different
$\beta$ for the model elliptical flow at either large or small $\mathrm{Ro}$.

\begin{figure}
\includegraphics[width=8cm]{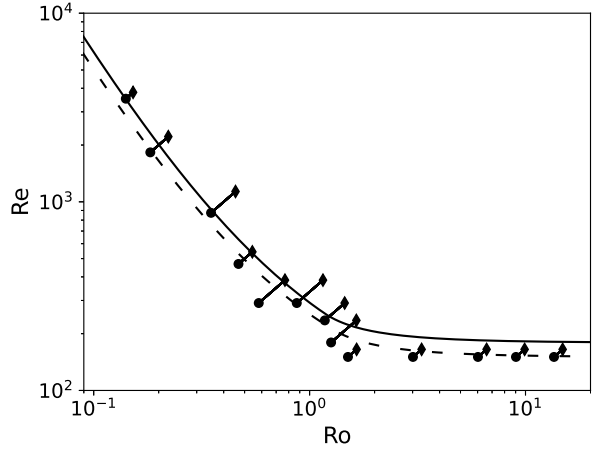}
\caption{
The stability limit of an elliptical vortex computed for 
$k_\parallel=5 \pi/3$, $k_\perp=2 \pi$ and $\beta=0.8$ (solid line,
intended as a fit at small $\mathrm{Ro}$) or
$\beta=0.76$ (dashed line, intended as a fit at large $\mathrm{Ro}$)
together with the same numerical data as in fig. \ref{fig:5}
delimiting the stability limit of 2D convection flow.}
\label{fig:8}
\end{figure}

In addition to using two different $\beta$, there are more refinements that are
possible to improve the stability line in fig. \ref{fig:5}. 
Visualisations of the velocity field just above onset computed in the previous
section show that the fastest growing mode prefers to fit a full wavelength into
the cell in the plane perpendicular to the rotation axis
so that we may set $k_\perp=2 \pi$. Furthermore, since the conditions
(\ref{eq:limit_k}) reproduce only approximately the effect of free slip
boundaries, we may consider $k_\parallel$ or $k_\perp$ to be fit parameters.
Leaving $k_\perp$ at $2\pi$ and slightly adjusting $k_\parallel$ to $5\pi/3$
leads to fig. \ref{fig:8}. While these adjustments of course improve the fit, it
should be noted that the straightforward choice of parameters in fig.
\ref{fig:5} suffices to reproduce the general appearance of the stability curve
and that the more refined choice only leads to minor modifications, separately
at large and small $\mathrm{Ro}$.

The values of $\left[ (a/b)^2-1 \right] / \left[ (a/b)^2+1 \right]$ in fig.
\ref{fig:6} are maximal around $\mathrm{Ro} \approx 1$ and it is precisely at
this $\mathrm{Ro}$ that the lines drawn in fig. \ref{fig:8} fail to reproduce
well the correct stability limit. The correlation between
the transition from 2D to 3D in the simulated convection flow and the onset of elliptical
instability deduced from $\left[ (a/b)^2-1 \right] / \left[ (a/b)^2+1 \right]$
supports the claim that this transition is due
to an elliptical instability of the convection roll.

It is crucial to account for a finite cell size through condition
(\ref{eq:limit_k}) to reproduce the correct stability limit. To gain insight
into the role of this condition, it is instructive to look at eqs.
(\ref{eq:k}, \ref{eq:u0}) in a limit in which asymptotically correct solutions
can be obtained analytically \cite{Craik89,Waleff90,Kerswe02}, which is the
double limit of $\beta \ll 1$ and $\omega \gg 1$. To perform the calculation,
$\bm u_0$ is expanded in a series in $\beta$ as
\begin{equation}
\bm u_0 = \left[ \bm w_0(t) + \beta \bm w_1(t) +{\cal O}(\beta^2) \right]
e^{\sigma t}
.
\end{equation}
This expansion is inserted into the energy equation derived from eq.
(\ref{eq:u0}) by taking the dot product of this equation with $\bm u_0$:
\begin{equation}
\frac{1}{2} \frac{d}{d t} \bm u_0^2
=
-\bm u_0 (\bm A + 2 \bm \omega) \bm u_0
- \eta |\bm k|^2 |\bm u_0|^2
.
\label{eq:energy}
\end{equation}
The problem for $\beta=0$ is solved by
\begin{equation}
\bm w_0 = \frac{1}{2} e^{-\eta k_0^2 t}
\begin{pmatrix} 
-(1-\cos \theta) \sin(\Gamma_+ t + \phi) - (1+\cos \theta) \sin(\Gamma_- t + \phi) \\ 
-2 \sin \theta \sin(\Gamma t + \phi) \\ 
-(1-\cos \theta) \cos(\Gamma_+ t + \phi) + (1+\cos \theta) \cos(\Gamma_- t + \phi)
\end{pmatrix}
\end{equation}
where $\phi$ is an arbitrary phase and $\Gamma_\pm = \Gamma \pm 1$ with
\begin{equation}
\Gamma = 2 (1+\omega) \cos \theta
.
\label{eq:Gamma}
\end{equation}
Inserting this expression into the right hand
side of eq. (\ref{eq:energy}) yields to leading order in $\beta$:
\begin{equation}
\sigma=\frac{\beta}{4} \left[
(1-\cos \theta)^2 \sin[2(\Gamma_+ t + \phi)]
- (1+\cos \theta)^2 \sin[2(\Gamma_- t + \phi)]
-2(1-\cos^2 \theta) \sin(2t) \right]
-\eta k_0^2
.
\end{equation}
The average of $\sigma$ over long time intervals vanishes unless $\Gamma_+$ or
$\Gamma_-$ is zero, in which case the maximal growth rate is
\begin{equation}
\sigma=\frac{\beta}{16} \frac{(3+2 \omega)^2}{(1+\omega)^2}
-\eta k_0^2
.
\label{eq:sigma_max}
\end{equation}
This growth rate in turn is maximized by the smallest admissible $k_0^2$. In an
infinitely extended vortex, $k_0^2$ could be arbitrarily small and viscosity
would be irrelevant for the stability of the vortex. However, $k_0^2$ is limited
in the present application by either the boundaries or the finite size of the
convection rolls. If we naively require $k_0^2 > k_{\mathrm{min}}^2$, a positive
growth rate is obtained in eq. (\ref{eq:sigma_max}) for
$\eta < \beta/(2 k_{\mathrm{min}})^2$ at large $\omega$, which corresponds
to a fixed critical Reynolds number at small Rossby numbers and which is not at
all the behaviour observed in the simulations. On the other hand, the more
detailed condition (\ref{eq:limit_k}) with $k_\parallel=2\pi$ imposes
$k_0^2 \ge 4 \pi^2 / \cos^2 \theta$, which because of eq. (\ref{eq:Gamma}) and
either $\Gamma_+$ or $\Gamma_-$ being zero leads to
$k_0^2 \ge 16 \pi^2 (1+\omega)^2$. Inserting this expression into eq.
(\ref{eq:sigma_max}) and setting $\sigma=0$ to find the stability limit, we
obtain
\begin{equation}
\frac{\beta}{16} \frac{(3+2 \omega)^2}{(1+\omega)^2}
=
16 \pi^2 \eta (1+\omega)^2
\label{eq:sigma=0}
\end{equation}
which in the limit of large $\omega$ is equivalent to
\begin{equation}
\frac{1}{\eta} =
\frac{64}{\beta} \pi^2 \omega^2
.
\end{equation}
Translated to Rossby and Reynolds numbers, this relation implies that 
$\mathrm{Re} \propto \mathrm{Ro}^{-2}$
at the onset of instability which is close to what is observed in the
simulations. Apparantly, it is the limited extent of the convection roll along
its axis which is responsible for the right hand side in eq. (\ref{eq:sigma=0})
and the observed variation of the Rossby number at the transition.

One can also deduce from this derivation why the flow can become 3D despite
$\mathrm{Ro}$ being small. A small $\mathrm{Ro}$ guarantees the advection term
to be small compared with the Coriolis term in the Navier-Stokes equation if
both terms are computed for the base flow. The wavenumber introduced by the
instability, $k_0$, behaves for large $\omega$ as $k_0 \sim 4 \pi \omega$. The
magnitude of the second term in eq. (\ref{eq:u'}), 
$(\bm U \cdot \nabla) \bm u'$, is then proportional to $|\bm u'| \omega$ and
hence comparable to the Coriolis term in eq. (\ref{eq:u'}),
$2 \omega \bm{\hat y} \times \bm u'$. Similarly, the time derivative 
$\frac{\partial}{\partial t} \bm u'$ is also comparable to the Coriolis term in
this case. The geostrophic balance, which requires that only the pressure
gradient be comparable to the Coriolis term in eq. (\ref{eq:u'}) and which implies
2D flows, is thus broken
because the instability occurs on a length scale which varies as the
inverse of $\omega$.

\section{Conclusion}
This study was triggered by the advent of experiments which use centrifugal
acceleration to create a large effective gravity inside a convection cell. The
rotation also favors 2D flow structures. Global quantities like the heat flux
scale differently in 2D and 3D flows so that it becomes important to know under
which conditions the 2D convection is replaced by 3D convection.

A comparison of time scales associated with rotation and buoyancy suggests that
buoyancy is irrelevant to the transition from 2D to 3D flows. However, this
transition does not occur at a specific Rossby number independent of other
control parameters. The transition is due to an elliptic instability of the
convection rolls. This confirms that buoyancy is irrelevant to the transition.
It also implies that a finite viscosity is not necessary for the instability
since the elliptical instability is an inertial instability. On the other hand,
viscosity together with a finite extent of the convection rolls leads to a
Reynolds number dependent Rossby number at the transition. In the simulations
and in the experiment by \citet{Menaut19} the convection rolls are confined by
sidewalls. However, even in an annular geometry, the lateral roll size is finite
and is for example limited by the critical wavenumber near the onset so that the
volume of fluid exposed to the basic elliptical flow is restricted. According to
an analytic calculation valid in the limit of small ellipticity and Rossby
number, it is at any rate the size restriction along the roll axis which is
essential and which leads to the relation 
$\mathrm{Re} \propto \mathrm{Ro}^{-2}$
at the transition.

The experiments performed so far do not provide us with enough information about
the velocity field to conclusively test the scenario of the elliptical
instability. Note that the Rossby number in \citet{Jiang20} is based on the
computed free fall velocity and not on an actual measured velocity. The
comparison of simulations and experiments in the 
$(\mathrm{Ra},\mathrm{Ek})-$plane is not helpful because it depends on
the $\mathrm{Re}(\mathrm{Ra})$ relation which is affected by the cylindrical
geometry in the experiment by \citet{Jiang20} and by compressibility in the
experiment by \citet{Menaut19}. However, compressibiliy is not expected to
modify the stability criterion in the $(\mathrm{Re},\mathrm{Ro})-$plane because
it was found that the criterion for elliptical instability is independent of
compressibility in spheroidal geometry \cite{Clausen14}. Measurements of
temperature and pressure fluctuations allow \citet{Menaut19} to determine which
of their flows are 2D. These authors estimate velocity assuming either a balance between
pressure gradient and Coriolis acceleration or between pressure gradient and the
advection term in the momentum equation and find Rossby numbers of $0.01$ and
$0.1$, respectively. Starting from the Rossby number deduced from the assumption
of a geostrophic equilibrium and Ekman numbers in the range
$10^{-7}-10^{-6}$ typical of the experiments, one obtains Reynolds numbers 
in between $10^4$ and $10^5$, which means that
$\mathrm{Re} < 100 \mathrm{Ro}^{-2}$ is obeyed. The stability limit of the
numerical simulations at low Rossby numbers is approximately given by
$\mathrm{Re} \approx 100 \mathrm{Ro}^{-2}$. In as far as a quantitative
comparison is possible, theory and experiment agree on the prevalence of 2D
flows in nearly geostrophic equilibrium in the experiments by \citet{Menaut19}.

\begin{acknowledgments}
The authors acknowledge the help by Marie-Christine Volk who computed the 2D results
shown in figs. \ref{fig:1}, \ref{fig:2} and \ref{fig:6} as part of her
Bachelor's thesis. This work was funded
by the Deutsche Forschungsgemeinschaft (DFG) under the grant Ti 243/13.
\end{acknowledgments}


%

\end{document}